\documentclass[journal,transmag]{IEEEtran}

\IEEEoverridecommandlockouts                                                                  
\usepackage{mleftright}
\usepackage{algorithm}
\usepackage{pbox}
\usepackage{algpseudocode}
\usepackage{setspace,epsfig,psfig,amssymb,amsfonts,arydshln,graphicx,subeqnarray,amsmath,graphicx,epstopdf, xcolor, comment, multirow, cite,multicol}
\usepackage{varioref}
\usepackage{wrapfig}
\usepackage{enumitem}
\usepackage{dblfloatfix} 
\usepackage{gensymb}
\usepackage{textcomp}
\usepackage[export]{adjustbox}
\usepackage{fancyvrb}
\usepackage{breqn}
\usepackage{psfrag}
\usepackage{subcaption}
\usepackage{physics, color}
\usepackage{stackengine}
\usepackage{scalerel,amssymb,mathtools}

\def\delequal{\mathrel{\ensurestackMath{\stackon[1pt]{=}{\scriptstyle\Delta}}}}
\usepackage{fancyhdr}

\fancypagestyle{FirstPage}{
\lfoot{ \footnotesize{\copyright2022 IEEE.  Personal use of this material is permitted.  Permission from IEEE must be obtained for all other uses, in any current or future media, including reprinting/republishing this material for advertising or promotional purposes, creating new collective works, for resale or redistribution to servers or lists, or reuse of any copyrighted component of this work in other works. doi: 10.1109/TMECH.2022.3175949} }
}

\begin{document}
\title{
A Learn-and-Control Strategy for Jet-Based Additive Manufacturing
}

\author{ Uduak Inyang-Udoh, Alvin Chen and Sandipan Mishra
\thanks{This work was supported in part by the NSF Data-Driven Cyberphysical Systems Award \#1645648 and by the State of
New York ESD/NYSTAR program.}
\thanks{Uduak Inyang-Udoh was previously with, and Alvin Chen and Sandipan Mishra are currently with, the Mechanical, Aerospace and Nuclear Engineering
Department, Rensselaer Polytechnic Institute, Troy, NY 12180 USA {\tt\small uinyangu@purdue.edu},  {\tt\small  chena17@rpi.edu}, {\tt\small  mishrs2@rpi.edu} }
}

\markboth{Preprint Submitted to IEEE Transactions on Mechatronics}{Bare Demo of IEEEtran.cls for IEEE Journals}

\maketitle


\begin{abstract}
In this paper, we develop a predictive geometry control framework for jet-based additive manufacturing (AM) based on a physics-guided recurrent neural network (RNN) model. Because of its physically interpretable architecture, the model’s parameters are obtained by training the network through back propagation using input-output data from a small number of layers. Moreover, we demonstrate that the model can be dually expressed such that the layer droplet input pattern for (each layer of) the part to be fabricated now becomes the network parameter to be learned by back-propagation. This approach is applied for feedforward predictive control in which the network parameters are learned offline from previous data and the control input pattern for all layers to be printed is synthesized. Sufficient conditions for the predictive controller’s stability are then shown. Furthermore, we design an algorithm for efficiently implementing feedback predictive control in which the network parameters and input patterns (for the receding horizon) are learned online with no added lead time for computation. The feedforward control scheme is shown experimentally to improve the RMS reference tracking error by more than $\boldsymbol{30\%}$ over the state of the art. We also experimentally demonstrate that process uncertainties are compensated by the online learning and feedback control.
\end{abstract}

\section{Introduction} \label{sec:introduction}
\thispagestyle{FirstPage}
Jet-based AM refers to manufacturing techniques in which droplets of a material to be fabricated are deposited
onto a substrate based on a input pattern. These droplets then solidify (either by polymerization or solidification), creating a solid layer. As this is carried out layer after layer, the 3D part develops. Like other AM techniques, the additive nature of jet-based AM enables the efficient manufacture of intricate and miniature parts which are otherwise difficult to fabricate. Hence, jet-based AM techniques find application in the manufacture of electronics, medical models, and compliant features for robots and synthetic tissues \cite{Shirazi2015,Singh,Guo2017, Wang2017}.

A major objective in these applications is to ensure that the droplet deposition results in parts that conform to desired geometry. Several studies have focused on heuristically tuning fabrication parameters such as the rate of deposition, spacing between droplets, and temperature to find the suitable process parameters acceptable \cite{Qi2019}. However, these lack a formal relationship between the deposited droplets and the ensuing height profile, which is necessary to optimally control the finished geometry. Moreover, solely heuristic tuning precludes the ability to compensate for droplet or layer height uncertainties during the fabrication process.

The earliest work on 3D part geometry control for jet-based AM was reported in \cite{cohen2010geometric} where the authors implemented a so-called greedy geometry feedback scheme on parts made from paraffin wax. A clear drawback of the scheme is its heuristic control law and lack of a droplet deposition model. Subsequent approaches in geometry-level control have incorporated a model into the control scheme. In \cite{lu2015layer}, a stochastic greedy-type control algorithm was employed based on an empirical model. However, this model is not generalizable and suffers from poor scalability. Other researchers have used simplified linear models for control. In \cite{Hoelzle2016}, the deposited droplets are modelled as a Gaussian distribution and a spatial iterative learning control (SILC) scheme was proposed to minimize geometry tracking error. By design, the SILC is limited to feedforward, or feedback control only if the reference profiles for each layer are identical \cite{AARNOUDSE201997}. In \cite{guo2016predictive} the droplets were assumed to be hemispherical and a predictive feedback controller was proposed and validated in simulation. 

Practical implementation of geometry control in high resolution AM requires the process model to be fairly accurate for feedforward control and requires the layer control input patterns to be synthesized in a timely manner for feedback control \cite{Landers2020ARO}. \cite{Inyang-Udoh2020} experimentally demonstrated model predictive control (MPC) using the linear graph-based model presented in \cite{guo2018control}. Though the graph-based model is a substantial simplification of the actual height evolution dynamics, the work showed that by implementing feedback control in a layer-wise fashion, geometries can be accurately tracked, although with increased lead time for computations.

In this work, we use the physics-guided data-driven model presented in \cite{Inyang-Udoh2021}. The model is a convolutional RNN (convRNN) with a physically interpretable architecture. This model accurately predicts the height evolution of parts under various scenarios using sparse data. Because of the model's accuracy, we deploy it for feedforward control and show significantly improved feedforward control performance over the state-of-the-art in \cite{Inyang-Udoh2020}. We develop an adaptive control framework in which the model parameters may be updated online while the process is simultaneously controlled in a closed loop fashion. This is made possible due to the low data requirement of the convRNN. The online learning and control frame work follows a strategy presented in \cite{Inyang-Udoh}. In simulation, the model was learned after each layer and the control input pattern for the next layer was computed. By using the model developed in \cite{Inyang-Udoh2021}, we allow the learning to occur after an arbitrary number of layers and generalize a nonlinear MPC for an arbitrary number of prediction layers. Moreover, the algorithm is designed efficiently such that no computational lead time, other than that required for measurement, is added to the process and thus may be implemented in practice.

The paper is organized as follows. Table \ref{table:notation} summarizes relevant notations. In Sec. \ref{sec:prob_descptn} we describe the control problem. The following section briefly reviews the state of the art in geometry control in droplet based AM methods. In Sec. \ref{subsec:pred_contrl} we develop a predictive controller for the process. The dynamical stability of the predictive controller is analyzed in Sec. \ref{sec:pred_ctrl_stability}. Sec. \ref{sec:l&c} discusses a strategy for implementing predictive control in an efficient manner. In Sec. \ref{sec:exp_results}, we demonstrate both feedforward and feedback (with online learning) control. Sec. \ref{sec:conclusion} concludes the paper and previews future work.

\begin{table}[h!]
\centering
\begin{tabular}{|l c|} 
 \hline
 \textbf{Term} & \textbf{Notation} \\ [0.5ex]
 \hline
  
  Reference, measured, model height profile & $R$, $Y$, $\hat{Y}$\\
  Input pattern & $U$\\
  $R$, $Y$, $\hat{Y}$, $U$ vectorized & $r$, $y$, $\hat{y}$, $u$\\
  $U$, $u$ at ${i}^{\textrm{th}}$ layer & $U^{(i)}$, $u^{(i)}$ \\
  Set with elements from $i$ to $j$& $\mathcal{U}_{(i)}^{(j)}$\\
  Integers between $0$ and $N$  & $[\![0,  N]\!]$\\ 
  Catenation of vectors $u^{(L+i)},i \in [\![0,  Z-1]\!]$ & $U^L$\\
  ConvRNN model parameters & $\theta$\\

  State transition function & $\phi_1$\\
  Softplus function & $\phi_2$\\
  ConvRNN internal state at time $t$ & $h_t$\\
  Admissible input matrix at $t$ & $U_t$\\
  $U_t$ vectorized & $u_t$\\
  $i^{\textrm{th}}$ element of the vector $u_t$ & $u_t(i)$\\
  Total ConvRNN time steps for layer $L$ & $N_L$\\
  Initial ConvRNN time step for layer $L$ & $t_{0_L}$\\
  Convolution kernel & $b$\\
  Toeplitz matrix for $b$ at time $t$  & $W_{u,t}$\\
  Sparse Droplet Identifier Matrix & $I_t$\\ 
  $Q$ is a positive semi-definite matrix & $Q\succeq 0$\\
  $P - Q \succeq 0$  & $Q\preceq P$\\
  Vector of $n$ 0's & $\boldsymbol{0}_n$\\
  Matrix of $n$ by $n$ 0's & $\boldsymbol{0}_{n\times n}$\\
  Infinitesimal number& $\delta$\\
  Identity Matrix & $I$\\
  Total number of layers to be printed & $T_L$\\
  Prediction/control horizon & $Z$\\
  Layers implemented between measurements & $Z_u$\\
  Amount of input-output data pairs used & $\Delta\ell$\\[1ex]
 \hline
 \end{tabular}
 \caption{Table of relevant notations}
 \label{table:notation}
 \end{table}
 
\section{Problem Description}\label{sec:prob_descptn}
\begin{figure}
   \begin{center}
    \includegraphics[width=0.42\textwidth]{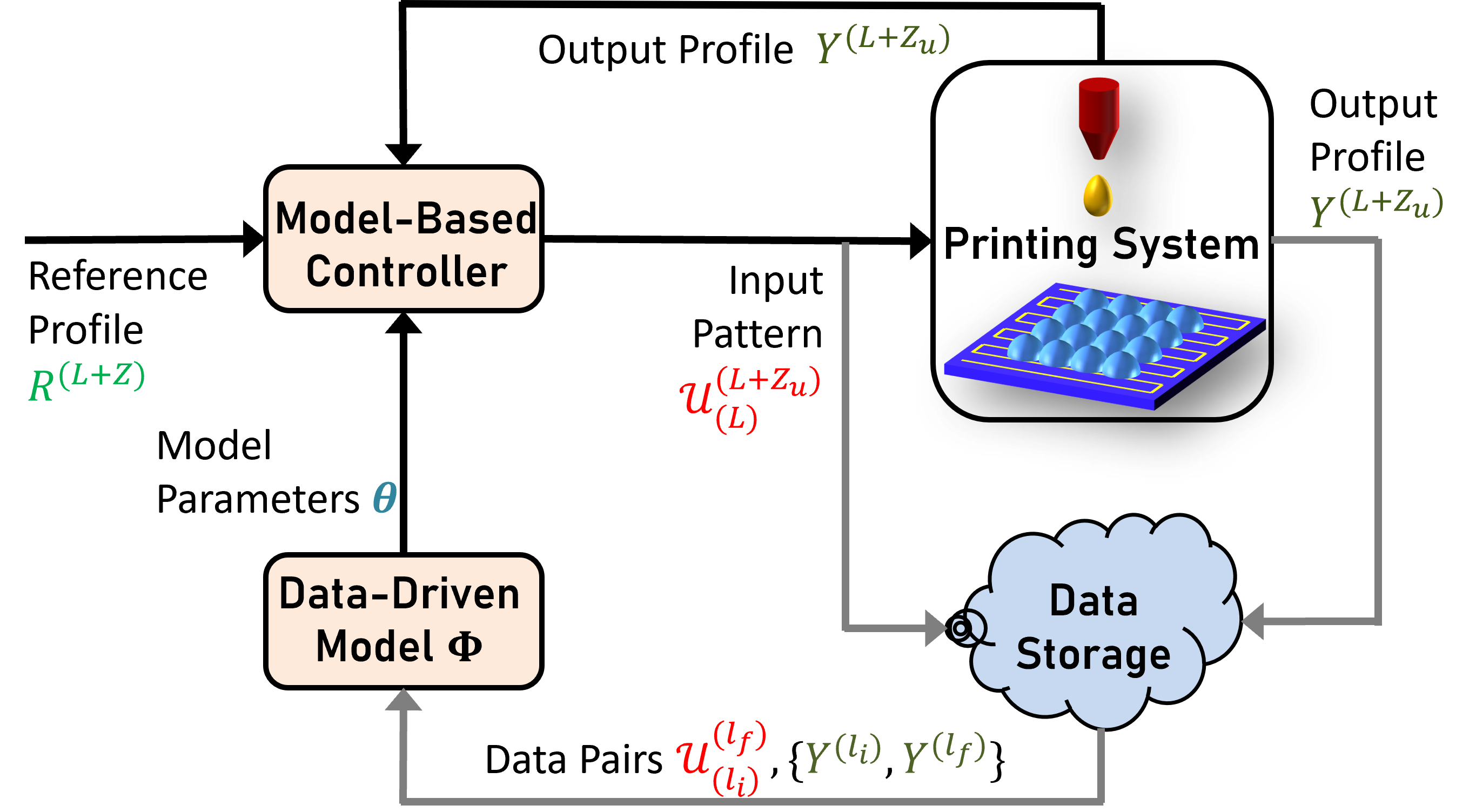}
   \caption{3D inkjet printing scheme. For each layer, a reference profile based on the desired part geometry is fed to a controller which generates a suitable input sequence.} 
      \label{pic:printer2}
   \end{center}
\end{figure}

Fig. \ref{pic:printer2} shows the basic scheme of the jet-based AM system considered in this work, a closed-loop drop-on-demand inkjet 3D printing process. A desired geometry $R$ is to be fabricated. The geometry is resolved in the horizontal plane into an $n_x$ by $n_y$ grid space to obtained a discretized height distribution (or profile). It is additionally sliced horizontally into layers. Assume we have just printed layer $L$ and the current height profile is $Y^{(L)} \in \mathbb{R}^{n_x \times n_y}$. Now supposing the desired height profile of the reference geometry at layer $L+Z$ ($Z > 0$) is $R^{(L+Z)}\in \mathbb{R}^{n_x \times n_y}$, we aim to determine the sequence of future input patterns $\mathcal{U}_{(L)}^{(L+Z)} = \{U^{(L)}, U^{(L+1)},\hdots, U^{(L+Z-1)} \}$, for $U^{({L+{k}})} \in \mathbb{R}^{n_x \times n_y}, {k} \in [\![0,  Z-1]\!] $, that achieve the desired height profile. Moreover, we desire to compensate for process uncertainties or change in the evolution of the height profile.  This calls for a feedback control framework in which the system's model may be adaptively updated in real-time. 

To design the control framework, we use the data-driven dynamical model presented in \cite{Inyang-Udoh2021} as it lends  itself to in-process learning. Consider the function $\Phi$ of \cite{Inyang-Udoh2021} parameterized on $\theta$ such that $\hat{Y} ^{(L+1)} = \Phi(\theta,{Y} ^{(L)},U^{(L)}) \in \mathbb{R}^{n_x \times n_y}$. The optimal set of  parameters $\theta^{*}$ is obtained by minimizing the error ${\|{Y}^{(\ell_f)}-\hat{Y} ^{(\ell_f)}(\theta)\|}_2^2$ using stored data pairs of $\mathcal{U}_{(\ell_i)}^{(\ell_f)} $ and $\{{Y}^{(\ell_i)}, {Y}^{(\ell_f)}\}$. ${Y}^{(\ell_i)}$ is the measured initial height profile at layer $\ell_i$, upon which the input sequence $\mathcal{U}_{(\ell_i)}^{(\ell_f)} $ produces ${Y}^{(\ell_f)}$ at layer $\ell_f$. Once $\theta^*$ is obtained, $\Phi$ may be written such that the control input becomes the optimization variable $\hat{Y} ^{(L+1)} = \Phi^*(\hat{Y} ^{(L)},U^{(L)})$. The optimal control input sequence for layers $L$ to $L+Z$, $\mathcal{U}_{(L)}^{(L+Z)*}$, may now be obtained by minimizing ${\|R^{(L+Z)}-\hat{Y} ^{(L+Z)}\|}_2^2$ using the pair ($Y^{(L)}$, $R^{(L+Z)}$).  To compensate for uncertainties after printing some layer $L+Z_u$ ($Z_u \leq Z$), feedback information (the height profile ${Y} ^{(L+Z_u)}$) may be collected and the control input for the next $Z$ layers may be recomputed. To alleviate plant-model mismatch, the set of parameters $\theta$ is updated using data pairs obtained of $\{{Y}^{(\ell_i)}, {Y}^{(\ell_f)}\}$ and $\mathcal{U}_{(\ell_i)}^{(\ell_f)} $ during the printing session. Subsequently, our two-fold objectives are: (1) to develop a procedure for finding optimal control sequence $\mathcal{U}_{(L)}^{(L+Z)*}$; and (2) to generate an efficient strategy for an online update of set $\theta^*$ and implementing feedback control.

\section{Related Work} \label{sec:past_models_and contrl}

In Sec. \ref{sec:introduction}, we note that several strategies have been employed  for geometry-level feedback control in jet-based AM processes. In this section, we summarily discuss these control strategies and assess the drawbacks that springboard the control framework presented in this work. 

\textit{Greedy Feedback Control:} Early attempts to implement geometry-level control involved using greedy algorithms to determine when and where to deposit droplets. In \cite{cohen2010geometric}, the next location to deposit a droplet was chosen based on a heuristic score assigned to each location. The algorithm is quite limited as it does not consider droplet interaction post-deposition or droplet size/scale. In \cite{lu2015layer} a `stochastic greedy-type' search algorithm is employed to minimize tracking error $\|R-\hat{Y}^{(L)}\|_2^2$ and surface roughness. This algorithm empirically models the effect of neighboring droplets when the edges of the part shrink. Consequently, it is not applicable in scenarios where the edges are elevated \cite{Inyang-Udoh2021}. Further, due to the `greedy' search approach employed, the algorithm increases in expense proportionally to the size of the grid resolution. 

\textit{Linear Time-Invariant MPC:} In \cite{Inyang-Udoh2020}, following the lifted model of \cite{Guocontrol2017}, the geometry control objective is formulated as an MPC problem, where a height-tracking cost-function $J$ is to be minimized over a finite receding horizon of $Z$ layers:
\begin{equation}\label{eq:cf}
\begin{aligned}
& \underset{U^{(L)}}{\text{min}}
& & J\left(U^{L}\right) \\
& \text{s.t.}
& & \hat{y}^{(L+k+1)} = \mathcal{A} \hat{y}^{(L+k)} + \mathcal{B} u^{(L+k)}, {k} \in [\![0,  Z-1]\!], \\
& & & U_{\text{low}} \leq {U^{L}}\leq U_{\text{high}},
\end{aligned}
\end{equation}
where $U^{L}=[u^{(L)^T}~...~u^{(L+Z+1)^T}]^T \in \mathbb{R}^{nZ}$, $u^{(L+k)}$ is the $i^{th}$ layer control input in the receding horizon, and $\hat{y}^{(L)}$ is the current height. $(\mathcal{A}, {\mathcal{B}})$ models the height evolution from layer to layer:  $\mathcal{A}\in \mathbb{R}^{n \times n}$ is the state  matrix  that  captures the dynamics of the height evolution over the entire layer; ${\mathcal{B}}\in \mathbb{R}^{n \times n}$ is the input matrix accounting for the height distribution of each droplet on deposition. The cost function is designed to penalize tracking error. $U_{\text{low}}$ and $U_{\text{high}}$ define the upper and lower constraints on the input. The optimization is performed each layer, and $u^{(L)^\star}$ is applied. The model $({\mathcal{A}}, {\mathcal{B}})$ does not adequately capture nonlinear fluid behavior when droplets overlap. Hence, the control approach requires layer-to-layer feedback to adequately compensate for the plant-model mismatch, in addition to compensating for uncertainties. Further, if operating conditions change, the model would require offline re-identification.

\textit{Iterative Learning Control:}
In \cite{Hoelzle2016}, a spatial iterative learning (SILC) scheme is proposed with the following learning law:
\begin{equation}
    u^{(L+1)} = \Gamma_uu^{(L)} + \Gamma_ee^{(L)}
\end{equation}
Here, $\Gamma_u, \Gamma_e \in\mathbb{R}^{n \times n}$ are the input and error learning matrices. $e^{(L)}$ is the error between the desired and model output: $e^{(L)} ={y}^{(L)} - \hat{y}^{(L)}$. $\hat{y}^{(L)}$ is evaluated in a similar way as in \eqref{eq:cf}, but with $\mathcal{A} = I$. As with the linear MPC strategy, $\Gamma_u, \Gamma_e$ are evaluated to minimize a height tracking cost function. The SILC is limited to feedforward, or feedback control \textit{if the reference profiles for each layer are identical} \cite{AARNOUDSE201997,Afkhami}. Though promising, typical demonstrations of the SILC scheme have been limited to simulations.

The above approaches have been shown to achieve improvement in geometry output \cite{Wang2018,Inyang-Udoh2020}. However, the model performance is limited as the linear models do not well capture the nonlinear fluid behavior when droplets interact. Furthermore, it is difficult to propagate any plant-model mismatch (disturbance) since the reference geometry for each layer is unique. Hence, it is important to use a model that: (1) more accurately captures the fluid behavior, (2) can be refined online if necessary, and (3) can be used to design a control strategy. Next, we employ the model presented in \cite{Inyang-Udoh2021} and design a predictive controller based on the model.  

\section{Predictive Control of the Printing Process}
\label{subsec:pred_contrl}
In the physics-guided convRNN model proposed in \cite{Inyang-Udoh2021}, the model (network) parameters are time invariant, while the input pattern is fed to the network as a time series. In this section, we re-express the model such that the input is now time-invariant. This re-expression allows us determine the (sub)optimal control input pattern by leveraging the similar gradient expressions as were used for the model identification.   

\subsection{ConvRNN Model Reformulated for Predictive Control}
In \cite{Inyang-Udoh2021}, for a given layer $L$, the time-step to time-step evolution of the height evolution was modeled as:
\begin{equation}\label{eq:h_kc1}
\begin{split}
&h_{t+1} = \phi_1(h_{t})+ \text{vec}~(b * U_{t}),  \hspace{.5cm}  t \in [\![0,  N_L-1]\!], \\
&\hat{y}^{(L+1)} = \phi_2(\phi_1(h_{N_L})), \hspace{.5cm} h_0 = \hat{y}^{(L)},
\end{split}
\end{equation}
where $h_t \in \mathbb{R}^{n}$, $ n = n_x \times n_y$, is the network's internal state (or height). The function $\phi_1 (h_{t})$ is defined as:
\begin{equation}\label{eq:h_kc2}
\phi_1 (h_{t}) =  h_{t}-D\sigma(\kappa D^Th_{t}), 
\end{equation}
and denoting $h^{(L)} \coloneqq  \phi_1(h_{N_{L-1}})$,  $\phi_2 (h^{(L)})$ is defined elementwise as:
\begin{equation}\label{eq:h_kc3}
 \phi_2(h^{(L)}(i)) =  \log(\gamma + \exp{h^{(L)}(i)+v_0}).
\end{equation}
In \eqref{eq:h_kc2}, $D \in \mathbb{R}^{n\times n_l}$ is an incidence matrix that  transforms the height profile vector into height differences across links where $n_l$ is the number of links. These differences are then weighted by a flowability constant $\kappa$ such that $\kappa D^Th_t$ is the effective flow across links at time $t$ due height to differences across each link, $D^Th_t$. The activation function $\sigma$ thresholds the effective height difference, $\alpha$, across a link that would cause flow at the time $t$. The soft-thresholding is applied to capture surface tension effect. (The reader is referred to \cite{Inyang-Udoh2021} for additional details). Meanwhile, $\phi_2$ in \eqref{eq:h_kc3} is a generic softplus function, parameterized on $\gamma$, that is applied to the element-wise sum of the internal state $h^{(L)}$ and a negative scalar $v_0$. The softplus function accounts for the curing effect and ensures the output profile $\hat{y}^{(L)}$ is non-negative.  

In \eqref{eq:h_kc1}, $b \in\mathbb{R}^{p \times p}$ is a convolution kernel that represents the effect of a droplet deposition on the height profile. $U_t \in\mathbb{R}^{n_x \times n_y}$, $t \in [0, N_L-1] $ are the admissible inputs at time step $t$ such that $\sum_{t=0}^{N_L-1} U_t = U^{(L)}$. The 2D convolution can be expressed as $\text{vec}(b * U_t) = W_uu_t$, where $W_u$ is a Toeplitz matrix corresponding to kernel $b$ and $u_t \in\mathbb{R}^n$ is $U_t$ vectorized. Thus, we can now rewrite  \eqref{eq:h_kc1} as:
\begin{equation}\label{eq:h_kc4}
h_{t+1} = \phi_1(h_t)+  W_{u,t}u^{(L)},  t \in [0, N_L-1] 
\end{equation}
Here, $u^{(L)}$ is the input vector for the entire layer and $W_{u,t} = W_uI_t$. $I_t$ is a sparse matrix holding a one at each position corresponding to where a deposition may take place: $I_tu^{(L)} = u_t$. Once the model parameters in \eqref{eq:h_kc1} are identified and given a reference profile, we can find a (sub)optimal input for layer $L$, $u^{(L)*}$ by gradient-based means.

\subsection{Predictive Control Using the Reformulated Model} \label{subsec:pred_control}
In \cite{Inyang-Udoh2021}, the optimal values of the model parameters were determined from input-output data. These values were determined via gradient descent direction using back-propagation through time. Now, we follow an analogous approach to determine (sub)optimal input, given knowledge of the model parameters and desired output. 
Assume layer $L$ has been printed and we are concerned with the output of the next $Z$ layers. We define cost function $J({U^{L}}) \delequal \sum_{i=0}^{Z} J^{(L+i)}$ where:
\begin{align}
J^{(L+i)} = \begin{cases} \norm{P (r^{(L+Z)}-\hat{y}^{(L+Z)})}_{2}^{2} & {i = Z},  \\ \norm{ Q (r^{(L+i)}-\hat{y}^{(L+i)})}_{2}^{2}  +  \norm{ G u^{(L+i)}}_{2}^{2}  &\textrm{else.}
\end{cases}\nonumber
\end{align}
The control optimization problem may be written as:
\begin{subequations}\label{eq:cntrl}
\begin{align}
{U^{L^*}} & = \underset{{U^{L}}}{\text{argmin }} 
J({U^{L}}) \\
\text{s.t. } h_{t_k+1} &= \phi_1(h_{t_k})+  W_{u,t_k}u^{(L)}, t_k \in [\![t_{0_k}, N_{L+k}-1]\!],\\
\hat{y}^{(L+k+1)} & = \phi_2(\phi_1(h_{N_{L+k}})), h_{t_{0_k}} = \hat{y}^{(L+k)}\\
\boldsymbol{0}_n & \leqslant{u^{(L+k)}}\leqslant \boldsymbol{1}_nu_{max} ~
\forall k \in [\![0, Z-1]\!]\,
\end{align}
\end{subequations}
where 
$P$, $Q$, $G$ $\in \mathbb{R}^{n\times n}$ are weighting matrices, and $u_{max}$ is an upper bound to the values in ${U^{L*}}$.
The total derivative of $J^{(L+i)}$ with respect to $U^{L}$ is:
\begin{align} \label{eq:tot_deriv}
&\dfrac{dJ^{(L+i)}}{{dU^{L}}} 
=  \pdv{J^{(L+i)}}{\hat{y}^{(L+i)}} \pdv{\hat{y}^{(L+i)}}{{U^{L}}} + \pdv{J^{(L+i)}}{{U^{L}}}\nonumber\\
&= 
\pdv{J^{(i)}}{\hat{y}^{(L+i)}} \sum\limits_{j=0}^{i} 
\sum\limits_{t_{j}=t_{0_{j}} }^{N_{j}} 
\pdv{\hat{y}^{(L+i)}}{h_{t_j}}  \pdv{h_{t_j}}{{U^{L}}} + \pdv{J^{(L+i)}}{{U^{L}}}. \quad
\end{align}
The terms in \eqref{eq:tot_deriv} may be expressed as:
\begin{align}
\pdv{J^{(L+i)}}{\hat{y}^{(L+i)}} &=  \begin{cases} 2({r^{(L+Z)}}-\hat{y}^{(L+Z)})^{T}P  ~\textrm{for } i = Z,\\
2({r^{(L+i)}}-\hat{y}^{(L+i)})^{T}Q ~\textrm{otherwise },
\end{cases}\\
\pdv{\hat{y}^{(L+i)}}{h_{t_j}} &= \pdv{\hat{y}^{(L+i)}}{h_{N_{L+i}}} \pdv{h_{N_{L+i}}}{\hat{y}^{(L+i-1)}} \cdots
 \pdv{\hat{y}^{(L+j)}}{h_{N_{L+j}}} \pdv{h_{N_{L+j}}}{h_{t_j}},\\
\pdv{h_{t_j}}{{U^{L}}} &= W_{t_j} = \begin{bmatrix}
  \boldsymbol{0}_{n\times n}^{(0)} \cdots \boldsymbol{0}_{n\times n}^{(i-1)}  W_{u,t_i} \cdots  \boldsymbol{0}_{n\times n}^{(Z-1)}
\end{bmatrix},\\
\pdv{J^{(L+i)}}{{U^{L}}} &=   \begin{cases}  0 &~\textrm{for } i = Z, \\u^{(i)^T}G &~\textrm{else; }
\end{cases} 
\end{align}
where:
\begin{align}
\pdv{\hat{y}^{({L+j})}}{h_{N_{L+j}}} &= diag(1/(1+\gamma \exp{\boldsymbol{1}v_0 - {h}_{N_{L+j}}})),\\
\pdv{h_{N_{L+j}}}{h_{t_j}} &= \prod_{N_{L+j}-1>k\geqslant {t_j}}\bigg( I - Ddiag (\sigma'(l_{k}))KD^T\bigg).
\end{align}
Then we can write that $\pdv{\hat{y}^{(L+i)}}{{U^{L}}} = \pdv{\hat{y}^{(L+i)}}{h_{N_{L+i}}} \pdv{h_{N_{L+i}}}{{U^{L}}}$ where:
\begin{equation}
\hspace{\dimexpr\displaywidth-\linewidth}
\pdv{h_{N_{L+i}}}{{U^{L}}} = \begin{cases}  0 ~ \textrm{for } i = 0, \\\sum\limits_{t_{1}=t_{0_{i_1}} }^{N_{L+1}} 
  \pdv{h_{N_{L+1}}}{h_{t_{1}}}  W_{t_{1}} ~ \textrm{for } i = 1,\\
 \sum\limits_{t_{i}=t_{0_{i}} }^{N_{L+i}}  \pdv{h_{N_{L+i}}}{h_{t_{i}}}  W_{t_{i}} + \pdv{h_{N_{L+i}}}{\hat{y}^{(L+i-1)}}  \pdv{\hat{y}^{(L+i-1)}}{{U^{L}}} ~  \textrm{else}; 
\end{cases}
\hspace{1000pt minus 1fil}
\end{equation}
Note that since the gradients are expressed analytically rather than as finite differences, the computation to find ${U^{L^*}}$ is expedited. We use the sequential quadratic programming method in which the Hessian is estimated using the Broyden–Fletcher–Goldfarb–Shanno algorithm and a penalty function is used to enforce the constraint \cite{fletcher_ch12,Bazaraa}. The approach is implemented using MATLAB's \texttt{fmincon} function.

\section{Predictive Control Stability}
\label{sec:pred_ctrl_stability}
In this section, we analyze the stability of the MPC problem. For this analysis, we neglect the  shrinkage effect in the output equation ($y = h_{N_L +1}$). We follow the Lyapunov's direct method \cite{chen1982receding}, where the cost function of the finite horizon optimization problem is used to establish stability \cite{keerthi1988optimal,rawlings1993stability,borrelli_bemporad_morari_2017}.

The height evolution from time step to time step within layer $L$ can be written as:
\begin{subequations}\label{eq:h_kc}
\begin{align}
&{h_{t+1} =  \phi_1 (h_{t}) + B_t\norm{u_{t}},  \hspace{.5cm}  t \in [\![0,  N_L-1]\!],} \\
&{\hat{y}^{(L+1)} = \phi_1 (h_{N_L}), h_0 = \hat{y}^{(L)},}
\end{align}
\end{subequations}
{where $B_t =  (1/ \norm{u_{t}})W_uu_t \in \mathbb{R}^{n}$},
\begin{equation}
\phi_1 (h_{t}) = (I -DK_{\delta}(h_t) D^T)h_{t} = A(h_{t})h_{t}
\end{equation}
{and $K_{\delta} (h_t) = diag(k_{\delta}) \in \mathbb{R}^{n_l \times n_l}$.} Each diagonal element is defined as:
\begin{equation}\label{eq:act_func1}
k_{\delta}(i) \delequal  \begin{cases}
				\kappa\big(1 - (1-\delta)\alpha/l_t(i)\big)  & \text{if  } l_t(i) > {\alpha},\\
				\delta \kappa & \text{if  } {-\alpha} \leqslant {l_t(i)} \leqslant {\alpha},\\
                \kappa\big(1 + (1-\delta)\alpha/l_t(i)\big) & \text{if  }   l_t(i) < {-\alpha},    
                \end{cases}
\end{equation}
with $l_t = \kappa D^Th_t \in \mathbb{R}^{n_l}$ and $\underline{\kappa} < \kappa < \bar{\kappa}$ ($\underline{\kappa}$ and $\bar{\kappa}$ are lower and upper bounds on the flowability constant). 
We can lift the time-step height evolution of \eqref{eq:h_kc1} for each layer to yield a layer-to-layer height evolution model to obtain $\hat{y}^{(L+1)} = \mathcal{A}_L\hat{y}^{(L)}+ \mathcal{B}_L u^{(L)}$
where $\mathcal{A}_L \in \mathbb{R}^{n\times{n}}$ is $\prod_{i=N_L}^{0}A(h_{i})$, $\mathcal{B}_L \in \mathbb{R}^{n\times{N_L}}$ is
\begin{align}
\label{Bprop}
\Big[\begin{matrix}
(\prod_{i=N_L}^{1}A(h_i))B_0 & (\prod_{i=N_L}^{2}A(h_i))B_1 &
\cdots 
\end{matrix} \nonumber\\
\begin{matrix}
(\prod_{i=N_L}^{t+1}A(h_i))B_t & (\prod_{i=N_L}^{t+2}A(h_i))B_{t+1}
\end{matrix}  \cdots &  A_{N_L}B_{N_L-1} \Big], \nonumber
\end{align}
{and we denote $\begin{bmatrix}
\norm{u_0} & \norm{u_1} & \cdots & \norm{u_t} & \cdots & \norm{u_{N_L -1}}
\end{bmatrix}^T \equiv u^{(L)}$.} Note that $\mathcal{A}_L$ and $\mathcal{B}_L$ are functions of $\hat{y}^{(L)}$ and $u^{(L)}$, and may be explicitly written as $\mathcal{A}(\hat{y}^{(L)},u^{(L)})$ and $\mathcal{B}(\hat{y}^{(L)},u^{(L)})$ respectively. For brevity, we retain earlier notations. 

Let error $e^{(L)} = r^{(L)}-\hat{y}^{(L)}$, where $r^{(L)}$ is the reference height profile for layer $L$. We assume there exists an ideal control input $u^{*(L)}$ such that $r^{(L+1)} = \mathcal{A}_Lr^{(L)} + \mathcal{B}_Lu^{*(L)}$. The closed-loop MPC system is now defined as  $e^{(L+1)} = \mathcal{A}_Le^{(L)} + \mathcal{B}_Lw^{(L)}$, where $w^{(L)} = u^{*(L)} - u^{(L)}$. The control objective is to minimize the following cost 
over the next $Z$ layers:
\begin{equation}\label{eq:cntrl2}
\begin{split}
& J({W_{L}})  \delequal (e^{({L+Z})})^T P e^{({L+Z})}
\\& \hspace{10pt} + \sum_{k=0}^{Z-1} (e^{({L+i})})^T Q e^{({L+i})} + (w^{({L+i})})^TG w^{({L+i})}.
\end{split}
\end{equation}

\noindent \textbf{\textit{Stability Lemma:}} The closed-loop MPC system $e^{(L+1)} = \mathcal{A}_Le^{(L)} + \mathcal{B}_Lf(e^{(L)})$, where $f(e^{(L)})$ is the receding horizon control law that associates the optimal input $w^{*(L)}$ to the current state $e^{(L)}$ is stable at the point $e^{(L)} = 0$ if:  
\begin{enumerate}
    \item $P = cP_D$ where $c$ is some constant and $P_D$ is a diagonal matrix such that $0 \preceq P_D \preceq I$, and
    \item $-P+Q+c\bar{A}^T\bar{A} \preceq 0$,
\end{enumerate}
where $Q {\succeq 0}$  and $\bar{A}=(I-\delta {\kappa} DD^T)$.  

\noindent \textbf{\textit{Proof:}} Define the Lyapunov function: $V(e^{(L)})=\underset{W_L}{\textrm{min}}J(W_L)$. Suppose the optimal input sequence is:
\begin{equation}
\begin{split}
W_L^{\star}(e^{(L)})&=\textrm{arg}\min_{W_L}J(W_L) \\ &=\{w^{*({L})},w^{*({L+1})},\cdots,w^{*({L+Z-1})}\}.
\end{split}
\end{equation}
The following shifted input sequence at layer $L+1$ is{:} 
\begin{equation}
\begin{split}
\tilde{W}_{{L+1}}(e^{({L+1})})=&\{w^{*({L+1})},\cdots, F_{{L+Z}}e^{({L+Z})}\},
\end{split}
\end{equation}
where $F_{L+Z}$ is some state feedback controller (gain) at $L+Z$. Note $\tilde{W}_{L+1}$ is not necessarily the \textit{optimal} input at layer $L+1$ for $e^{(L+1)}$.  Let $\tilde{V}(e_{L+1})=J(\tilde{W}_{L+1})$. We have:
\begin{equation} {\label{V_diff}}
\begin{split}
&\tilde{V}(e^{(L+1)})-V(e^{(L)})=-(e^{(L)})^TQe^{(L)}-(u^{(L)})^TGu^{(L)} \\&+(e^{(L+Z)})^T \Big(-P+Q
+(F_{L+Z})^TG(F_{L+Z}) \\&
(\mathcal{A}_{L+Z}+\mathcal{B}_{L+Z}F_{L+Z})^TP(\mathcal{A}_{L+Z}+\mathcal{B}_{L+Z}F_{L+Z})\Big)e^{(L+Z)},
\end{split}
\end{equation}
The first two terms on the right-hand-side of \eqref{V_diff} are non-positive. For the case where $F_{L+Z} = 0$, for any $L+Z$, we want to show that the third term is also non-positive. From Lemma Condition 1, it can be proved that
\begin{equation}\label{eq:P_ineq}
    \mathcal{A}_{L+Z}^TP\mathcal{A}_{L+Z}  \preceq c\bar{A}^T\bar{A} ~ \forall~{L+Z}.
\end{equation}
Since $-P+Q+c\bar{A}^T\bar{A} \preceq 0$ (Lemma Condition 2), we have:
\begin{equation}\label{eq:P_ineq2}
    -P+Q+\mathcal{A}_{L+Z}^TP\mathcal{A}_{L+Z}  \preceq -P+Q+c\bar{A}^T\bar{A}  \preceq 0 ~ \forall{L+Z}. \nonumber
\end{equation}
Thus, $\tilde{V}(e^{(L+1)})-V(e^{(L)}) \leq 0 ~ \forall ~ e_L \neq 0$. Now, noting that $\tilde{W}_{L+1}(e^{L+1})$ (for the case $F_{L+Z} = 0$) is not necessarily optimal, it follows that:
\begin{equation}
    {V}(e^{(L+1)})-V(e^{(L)}) \leq \tilde{V}(e^{(L+1)})-V(e^{(L)}) \leq 0 ~ \forall ~ e_L.
\end{equation}
\noindent\textit{Proof of Equation \eqref{eq:P_ineq}:}
We want to show that $ \mathcal{A}_{L}^TP\mathcal{A}_{L} \preceq \bar{A}^T\bar{A}$ $\forall L$.
Let ${A}_t= I -DK_t D^T$, where $K_t = K_\delta (h_t)$ as already defined. Recall that $P=cP_D$, where $P_D \preceq I$ (Lemma Condition 1), and $\bar{A} = (I-\delta {\kappa} DD^T)$. The inequality may then be written as:
\begin{equation}
\label{eq:proof_ineq}
    {A_{N_L}}\cdots{A_0}P_D{A_0}\cdots{A_{N_L}} \preceq \bar{A}^T\bar{A}
\end{equation}
Let $\mathcal{S}$ denote the set of all positive definite matrices with spectral radii less or equal to $1${,}  $\mathcal{S}=\{S : 0 \preceq S \preceq I; S =S^T\}$. To prove \eqref{eq:proof_ineq}, we need to show that for any element $M \in \mathcal{S}$, $A_t^TMA_t \in \mathcal{S} $ $\forall~ t$. 
\cite{Inyang-Udoh2021} {establishes} that the spectral radius of the Laplacian $DD^T$,  $\rho(DD^T) \leqslant 12$. Further, the lower and upper bounds on the flowability constant $\kappa$ {(}$\underline{\kappa}$ and $\bar{\kappa}${)} were given as $0$ and $1/6$ respectively. Since $K_t$ is diagonal, 
\begin{align}
    2(K_t - \delta \underline{\kappa} I)^{-1} \succeq 12I \succeq D^TD.
\end{align}
The second inequality follows from $\rho(DD^T) \leqslant 12$. Hence:
\begin{align}
    2(K_t - \delta \underline{\kappa} I) \succeq (K_t - \delta \underline{\kappa} I)D^TD(K_t - \delta \underline{\kappa} I),
\end{align}
which evaluates to:
\begin{align}
     I  - 2\delta \underline{\kappa} DD^T  & + (\delta \underline{\kappa})^2 DD^T DD^T \succeq \nonumber\\ & I - 2DK_tD^T + DK_tD^TDK_tD^T.
\end{align}
Thus, $\bar{A}^T\bar{A} \succeq A_t^TA_t$ for any $t$. By definition, $M \preceq I$, therefore:
\begin{align}\label{eq:atma}
    A_t^TMA_t \preceq A_t^TA_t\preceq \bar{A}^T\bar{A} \preceq I ~~ \forall t.
\end{align}

Furthermore, for any $M \in \mathcal{S}$, $M \succeq 0$ and so $A_t^TMA_t \succeq 0$; thus \eqref{eq:atma} can be expanded as:
\begin{align}\label{eq:atma2}
    0 \preceq A_t^TMA_t \preceq A_t^TA_t\preceq \bar{A}^T\bar{A} \preceq I ~~ \forall t.
\end{align}
Finally, \eqref{eq:atma2} implies that $A_t^TMA_t \in \mathcal{S}$ $\forall t$. {This result indicates that to guarantee stability, $P_D$ should be chosen such that $P_D\succeq \bar{A}^T\bar{A}$, and then, $0\preceq Q \preceq P-c\bar{A}^T\bar{A}$.}

\section{\textit{Learn \& Control:} Feedback Control Strategy} \label{sec:l&c}
\begin{figure}[h!]
\begin{center}
\includegraphics[trim=0cm 0cm 1cm 0.5cm, width=0.4\textwidth]{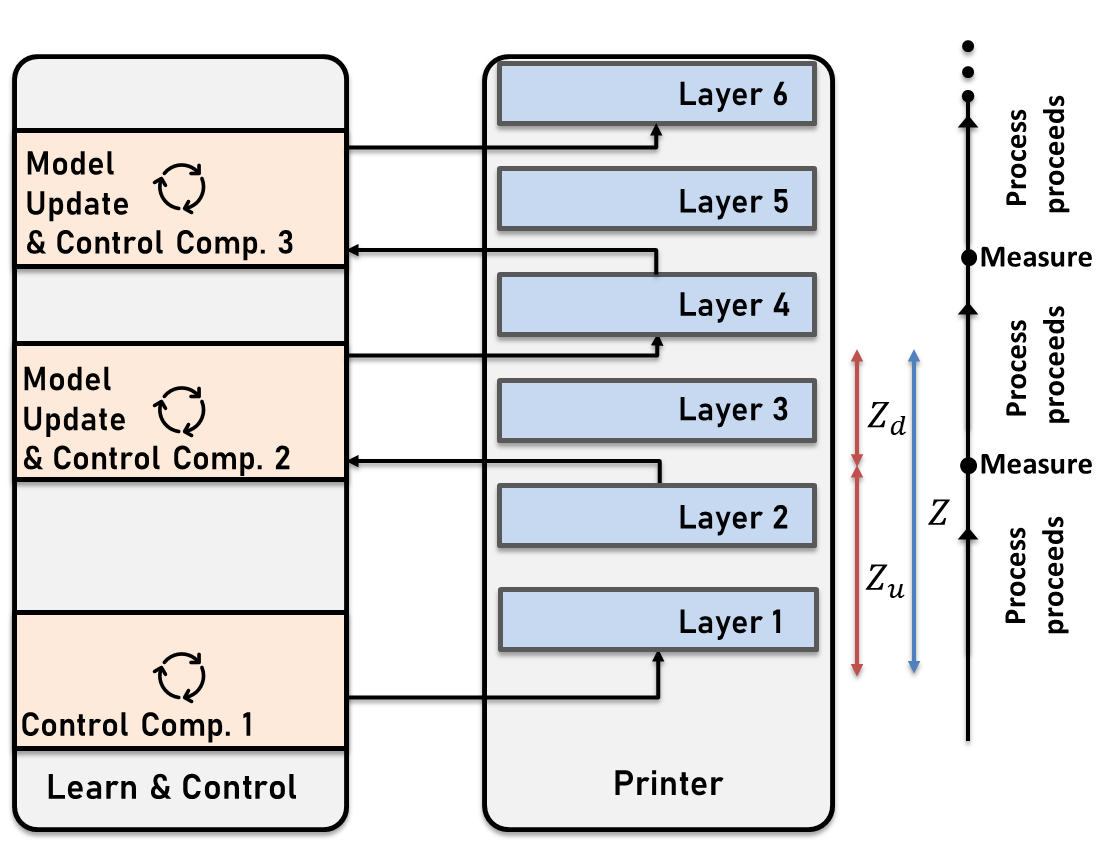}
\caption{{Illustration of feedback MPC strategy to eliminate downtime for online model learning and control computation.}}\label{pic:semi_mpc}
\end{center}
\end{figure}

\begin{figure*}[h!]
\begin{subfigure}[b]{0.7\textwidth}
   \begin{center}
   \includegraphics[trim=0cm 0.0cm 0.0cm 0.04cm,width=.97\textwidth]{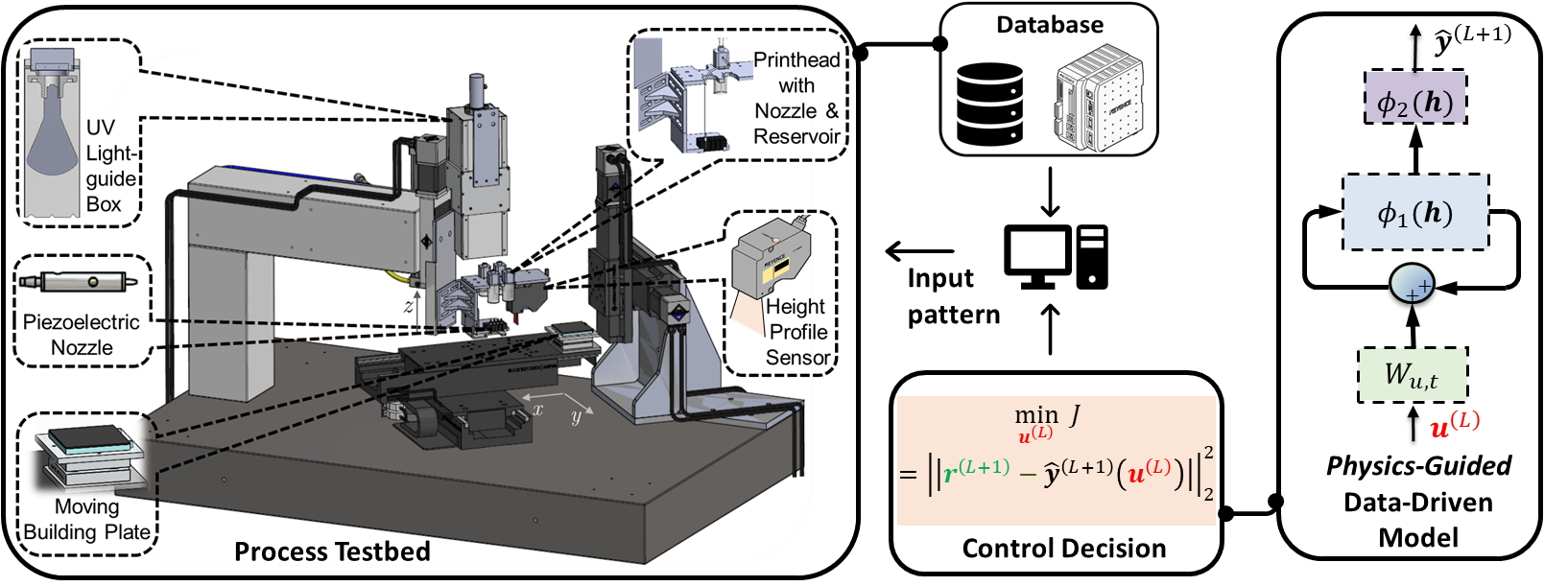}
   \caption{}
   \label{pic:ctrl_experiment}
   \end{center}
\end{subfigure}
\begin{subfigure}[b]{0.3\textwidth}
   \begin{center}
      \includegraphics[trim=0cm 0cm 0cm 0cm, width=1\textwidth]{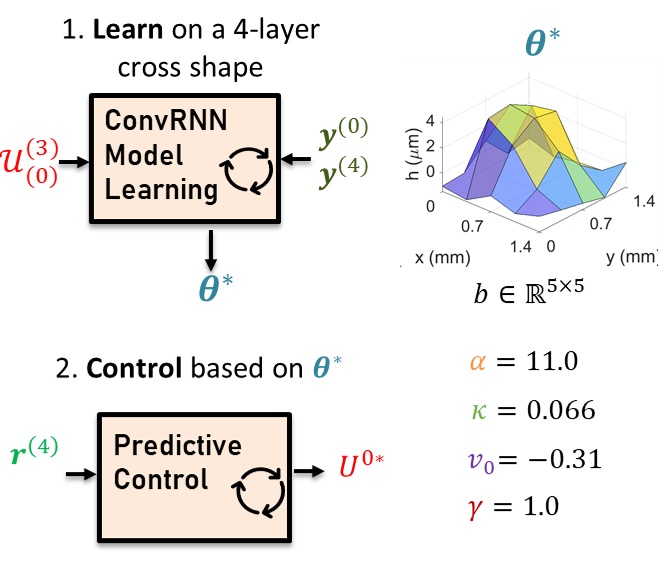}
\caption{}
   \label{pic:feedfor_exp}
   \end{center}
\end{subfigure}
\caption[8pt]{(a) Schematic of the control execution. {The} input pattern for each layer is generated using the convRNN-model-based controller. The computer directs the motion stages and nozzle droplet deposition based on the input pattern. The part is cured under UV light after each layer and the height is measured and stored as data for model training and control. (b) { Feedforward control. The model parameters are learned from a printed cross shape; these parameters are now used for MPC of other parts. The parameters of $\theta^*$ for the cross shape are given on the right.}}\label{pic:expt}
\end{figure*}

Given an accurate model of the height evolution, the optimal control scheme of Section \ref{subsec:pred_control} can provide good tracking performance. As the number of layers increases {however}, uncertainty in the printing process may begin to substantially impact the height profile evolution. In addition, given that the MPC depends on a data-driven model, we may want to update the model online utilizing data from the current print session (especially if  geometry or printing conditions change). {The} computational expense of updating (training) the model and computing new control input may lead to substantial lead times in the fabrication process. Therefore, in this section, we discuss a strategy for efficient adaptive feedback control. We propose a \textit{semi-feedback} approach that allows printing and computation to {occur} simultaneously. This strategy considers that the actual process is relatively slow and {that} necessary computations will be made in parallel with the printing of one or more layers. 

We begin {by} implementing the MPC similar to the traditional fashion, that is, we calculate input for $Z$ layers in the horizon, implement $Z_u$ layer(s) and obtain feedback; but while recomputing control input for the next $Z$ layers, we proceed to print $Z_d$ layer(s). 
The algorithm is demonstrated in Fig. \ref{pic:semi_mpc} for a 6-layer part and proceeds as follows:

\begin{enumerate} 
    \item Given the total number of layers to be printed $T_L$ and control horizon $Z$, set the data size to be used for training as $\Delta\ell$. Also select the number of layers $Z_u$ that {will} be implemented before {obtaining} the next feedback measurement, such that $Z_u+ Z_d\leq Z$ and $Z_u> Z_d$. Let $L = 0$ and define $T_s \delequal \left\lfloor(T_L - (Z_d+Z_u))/Z_u)\right\rfloor $.
    
    \item Use the last $\Delta\ell$ input layer pairs in the data base to identify the {set of} model parameters. If $\Delta\ell = 0$, assume {the} parameters {are} based on linear superposition.
    \item Calculate the control input for $Z$ layers into the future.
    \item For {$i = {0}\cdots{T_s}$:}
    \begin{enumerate} 
        \item Implement control input for $Z_u$ layers into the future.
        
        \item Get feedback measurement of the height profile $L = L + Z_u$ and add to database.
        
        \item Proceed to print $Z_d$ more layers. {Simultaneously}, update model parameters using {the} $\Delta \ell$ last input layer pairs in the augmented data base and calculate {the} control input for the next $Z$ layers, fixing $\big[
        u^{(L)^T} \cdots u^{(L+Z_d-1)^T}
        \big]^T$ already being implemented.
        \item Set $L \gets L+Z_d$. If {$i =  1$}, set $~Z_u \gets Z_u - Z_d$.
 
    \end{enumerate}
    
\end{enumerate}

\section{Experimental Results}\label{sec:exp_results}

\begin{figure*}[h!]
\begin{subfigure}[b]{0.67\textwidth}
   \begin{center}
   \includegraphics[trim=0.7cm 0.04cm 0.5cm 0.04cm,width=.941\textwidth]{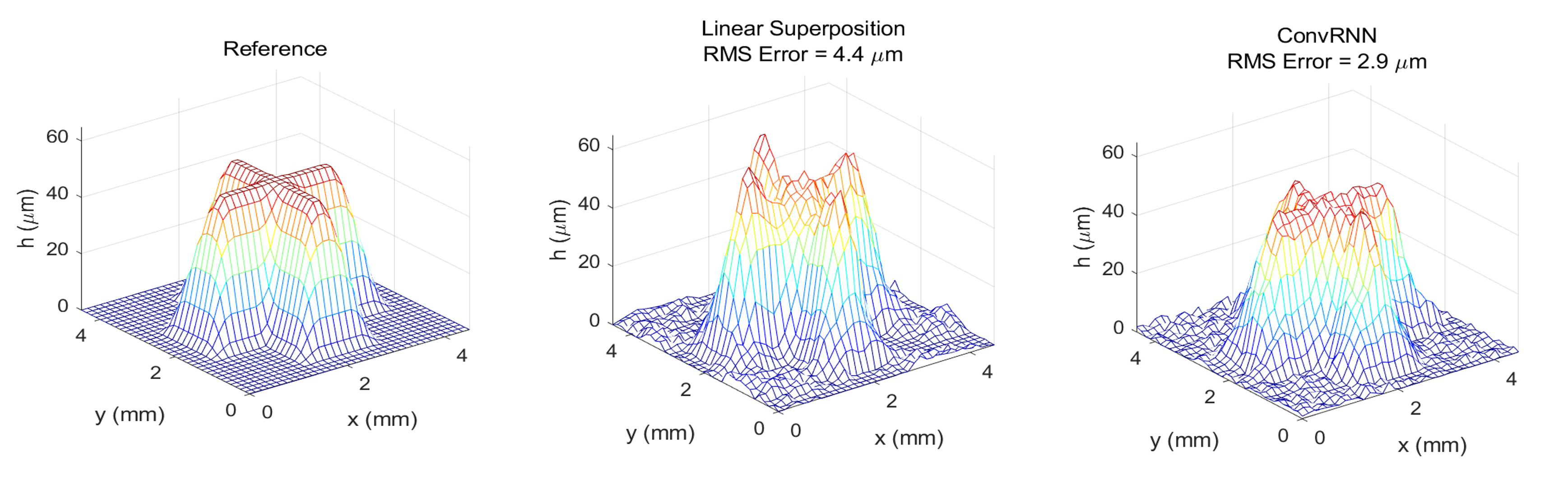}
   \caption{Height profile}
   \label{pic:ctrl_cross}
   \end{center}
\end{subfigure}
\begin{subfigure}[b]{0.3\textwidth}
   \begin{center}
      \includegraphics[trim=0cm 0cm 0cm 0cm, width=.951\textwidth]{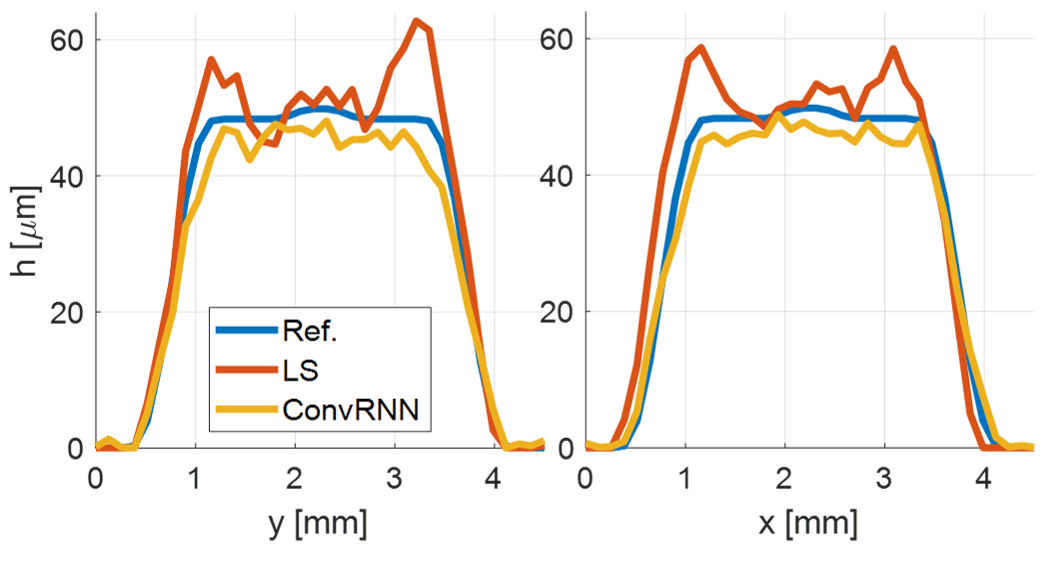}
\caption{Longitudinal section}
   \label{pic:ctrl_cross_sec}
   \end{center}
\end{subfigure}
\caption[8pt]{{Comparison of performance between the linear-superposition-based feedforward control and the convRNN-based feedforward control for a 4-layer cross-shape part. The measured grid resolution is $36 \times 36$}.}\label{pic:ctrl_cross1}
\end{figure*}

\begin{figure*}[h!]
\begin{subfigure}[b]{0.67\textwidth}
   \begin{center}
   \includegraphics[trim=0.7cm 0.04cm 0.5cm 0.04cm,width=.941\textwidth]{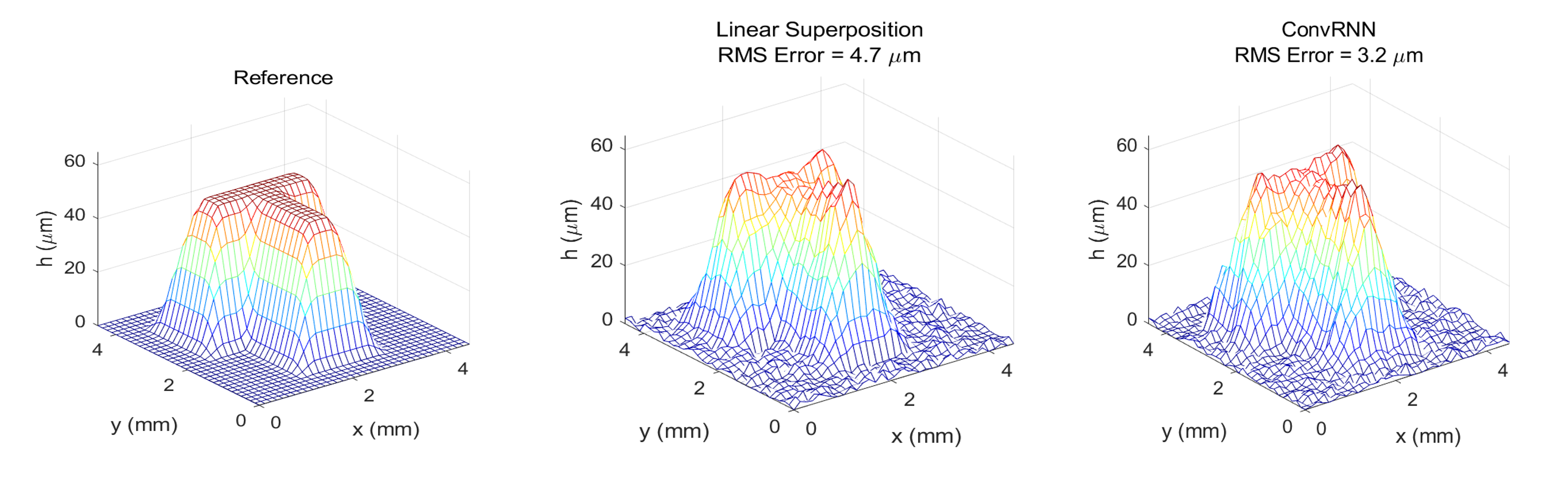}
   \caption{Height profile}
   \label{pic:ctrl_T}
   \end{center}
\end{subfigure}
\begin{subfigure}[b]{0.3\textwidth}
   \begin{center}
      \includegraphics[trim=0cm 0cm 0cm 0cm, width=.951\textwidth]{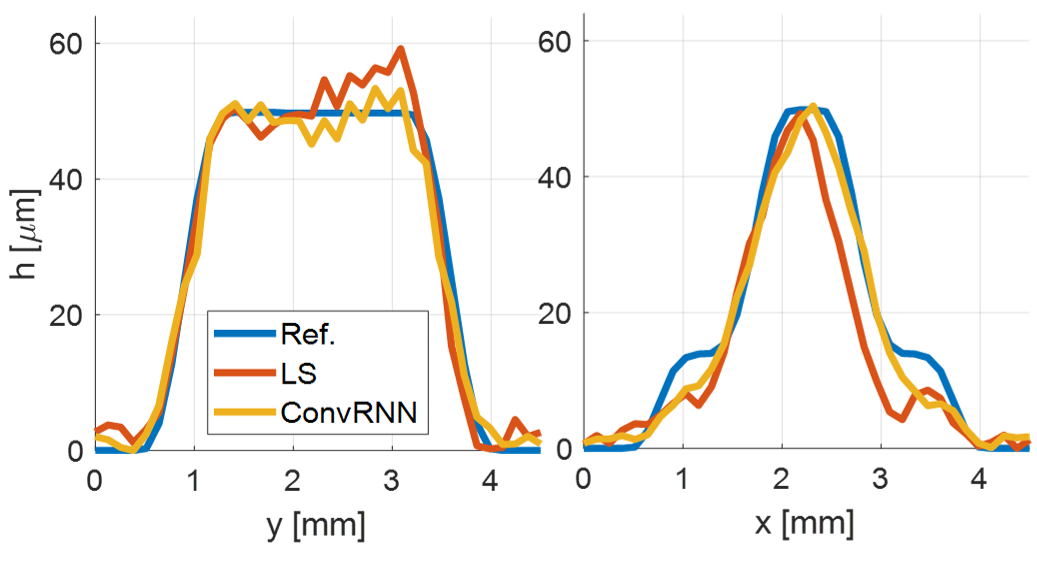}
\caption{Longitudinal section}
   \label{pic:ctrl_T_sec}
   \end{center}
\end{subfigure}
\caption[8pt]{{Comparison of performance between the linear-superposition-based feedforward control and the convRNN-based feedforward control for a 4-layer T-shape part. The measured grid resolution is $36 \times 36$}.}\label{pic:ctrl_cross2}
\end{figure*}

Fig. \ref{pic:ctrl_experiment} shows the experimental setup for control execution. {Initially}, the 3D model of {a} part is sliced horizontally into layers and an associated motion path for the nozzle is generated, along with droplet deposition locations. Motion stages move the build substrate while the nozzle deposits droplets according to the input pattern. {Once} all depositions for a layer are complete, the part is cured under UV light. Then {a} laser sensor measures the {layer} height profile for feedback control. The process is repeated until all layers are printed. 

\begin{table}[htb]
    \centering
    \begin{tabular}{l r}
        \hline
         Grid size ($n_x \times n_y$) & $18\times 18$ \\
         Grid spacing & 281.75$\mu$m\\
         Ink type  & Stratsys' TangoBlack FLX973 \\
         Substrate & Cured TangoBlack FLX973\\
         Volatge waveform\cite{microfab} & Bipolar (60V, 20$\mu$s dwell time)\\
         \hline
    \end{tabular}
    \caption{{Printing process parameters}}
    \label{tab:process_parameters}
\end{table}

\subsection{Feedforward Control Implementation}
\label{subsec:expt_feedforward}
In this subsection, we experimentally implement the feedforward MPC in Sec. \ref{subsec:pred_control}. {The driving signal applied to the piezoelectric nozzle (see \cite{microfab} for details), the ink type, substrate, and the grid dimensions used for the experiment are given in Table \ref{tab:process_parameters}. {We} print four layers of a cross shaped part}. 
{The convRNN model parameters are learned {from} this print using the single data pair, $\{{Y}^{(0)}, {Y}^{(4)}\}$ and  $\mathcal{U}_{(0)}^{(3)}$ (Fig. \ref{pic:feedfor_exp}). The identified parameters are given in Fig. \ref{pic:feedfor_exp}. First, we attempt to find a feedforward control input for a similar cross shape. We find the input pattern for each layer as specified in Sec \ref{subsec:pred_control}, solving for $U^0 \in \mathbb{R}^{4n}$ with $u_{max} = 2$. Because the system is limited to discrete droplets, the solution is quantized by rounding up to the nearest integer and implemented.  For comparison, a second cross-shape part with the same reference is printed based on the linear superposition model described in \cite{Guocontrol2017}. }The cross-shape parts printed based on the linear superposition and the convRNN models are meshed in Fig. \ref{pic:ctrl_cross}. {Note that the sensor measurement gives a finer resolution ($36 \times 36 $) than the above grid size.} The convRNN-based control yields a $34\%$ improvement in RMS tracking error over that of the linear superposition. Longitudinal sections through the parts {(}Fig. \ref{pic:ctrl_cross_sec}{) accentuate} the convRNN-based control compensation for elevation of the cross edges. 
We then carry out similar feedforward control for a 4-layered T-shape part (Fig. \ref{pic:ctrl_T}) using only the original identified model parameters $\theta^*$ from the cross shape. Similar RMS tracking error performance improvement is observed.  Fig. \ref{pic:ctrl_T_sec} shows  longitudinal sections through the T-shape part. {We observe} that the convRNN-based controller not only compensate{s} for elevation of the surface edges, but it also rectifies mismatch{es} in the side walls. An improvement of the reference tracking for all layers (not shown) is also noted. 

\begin{figure*}[h!]
\begin{subfigure}[b]{0.25\textwidth}
   \begin{center}
   \includegraphics[trim=1cm 0cm 0cm 0cm,width=.94\textwidth]{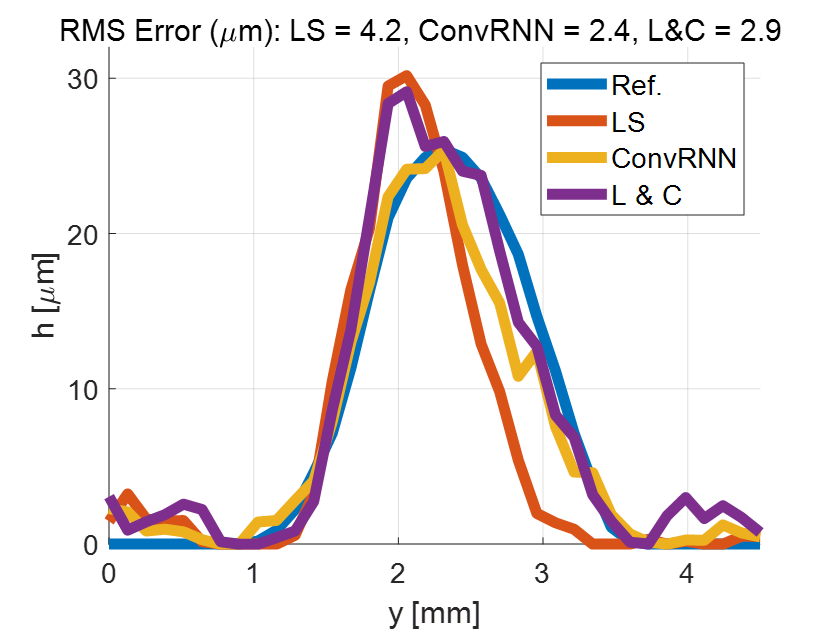}
   \caption{Layer 2}
   \label{fig:NN_printing_result_sectn1}
   \end{center}
\end{subfigure}
\begin{subfigure}[b]{0.25\textwidth}
   \begin{center}
      \includegraphics[trim=1cm 0cm 0cm 0cm, width=.94\textwidth]{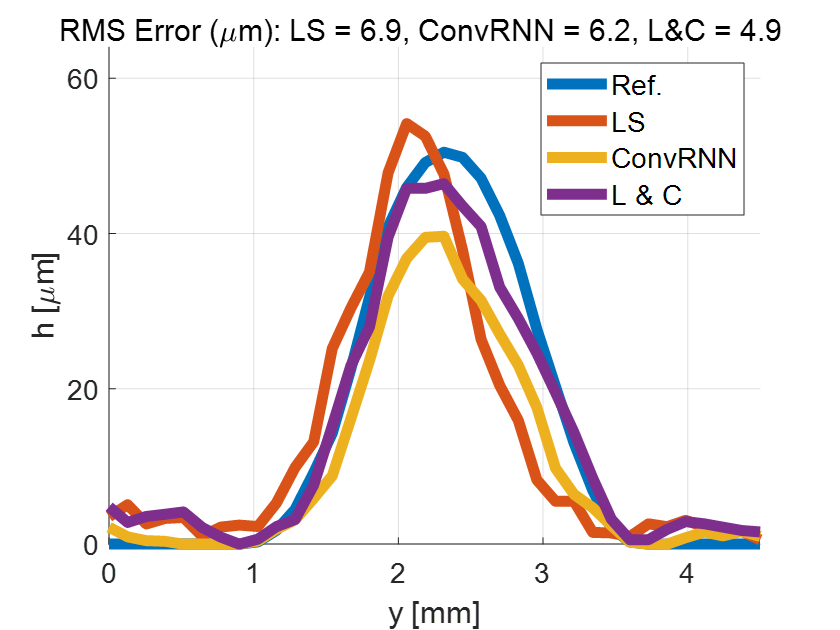}
\caption{Layer 4}
   \label{fig:NN_printing_result_sectn2}
   \end{center}
\end{subfigure}
\begin{subfigure}[b]{0.5\textwidth}
   \begin{center}
      \includegraphics[trim=3.5cm 0cm 0cm 2.5cm, width=1.0\textwidth]{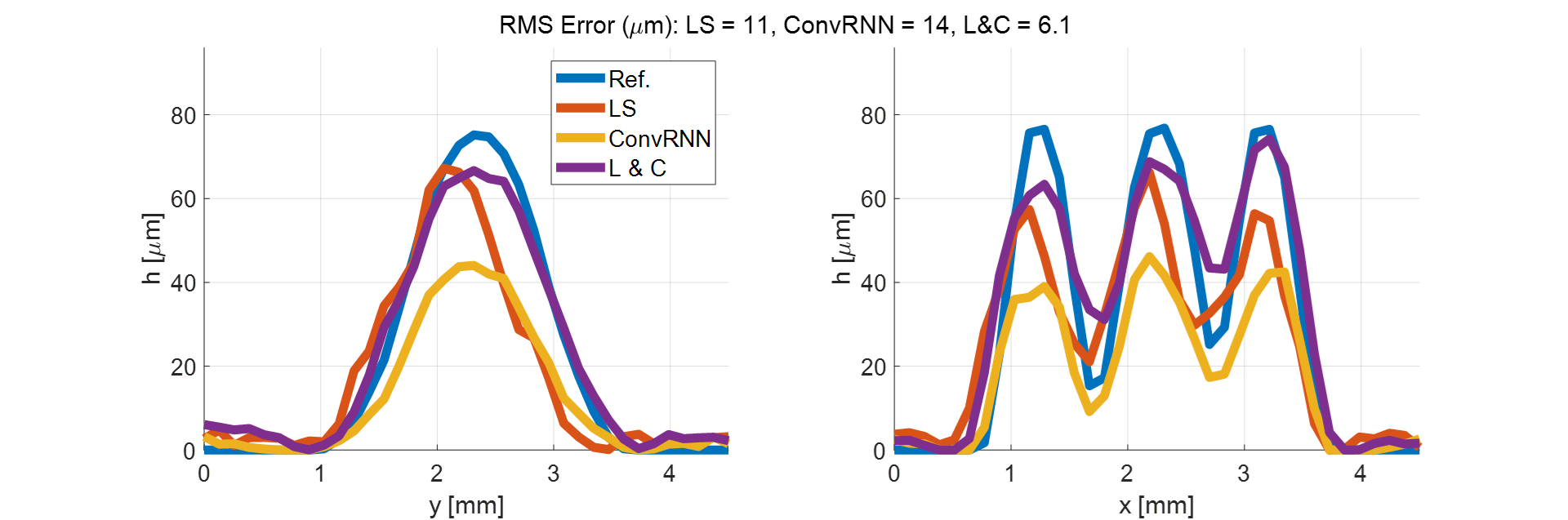}
\caption{Layer 6}
   \label{fig:NN_printing_result_sectn3}
   \end{center}
\end{subfigure}
\caption[8pt]{Longitudinal sections of open and closed loop (online learning \& control) profiles after every other layer of an N-shaped part.  At the top of each subplot is displayed the RMS error of each control approach with respect to the reference.}
\label{fig:learn_control}
\end{figure*}

\begin{figure}[h!]
\centering
\includegraphics[trim=0.7cm 0.04cm 0.5cm 0.2cm, width=0.36\textwidth]{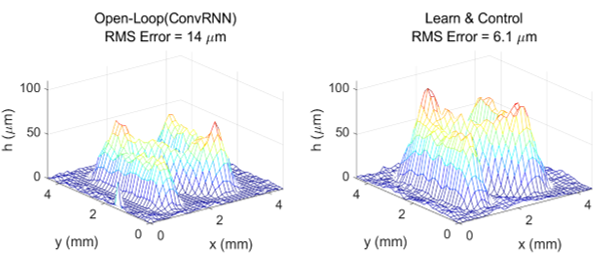}
\caption{Comparison of feedforward control performance of the convRNN with the learn \& control algorithm for an N-shape part. RMS error is based on the reference profile (not shown).}\label{pic:N_part}
\end{figure}

\subsection{Online Learning and Control Implementation}

We demonstrate the \emph{learn \& control}  strategy in Sec. \ref{sec:l&c} using an `N' shape printed in the same fashion as the parts in Sec. \ref{subsec:expt_feedforward}. We first print 6 layers of the part in open-loop based on the linear superposition model. {We print 6 additional layers of the part {using a} {feedforward} control input that relies on the identified convRNN model $\theta^*$ used in Sec. \ref{subsec:expt_feedforward}. Fig. \ref{pic:N_part} summarizes the results. The figure presents longitudinal sections through the printed part {every} other layer. Although the convRNN-based {feedforward} result outperforms that of the superposition at lower layers, as the number of layers grows, bias in the learned $\theta^*$ begin to dominate and convRNN-based {feedforward} is no longer advantageous.} Recall that the convRNN {feedforward} input {is based on merely one data pair} (the cross-shaped part in Sec. \ref{subsec:expt_feedforward}). Not only does the cross-shape part have fewer layers {than the N-shape part}, the shrinkage observed in printing this N-shaped part is more significant than for the {former}. Hence, it is imperative to \textit{populate} the dataset and learn from it as printing proceeds. Thus, we now print the same `N' shape with \emph{learn \& control} strategy developed in Sec. \ref{sec:l&c} as illustrated in Fig. \ref{pic:semi_mpc}. The following parameters are used: $Z = 3,~ Z_u = 2, ~Z_d = 1$.

Results for the online learning and feedback control strategy are superimposed on Fig. \ref{fig:learn_control} (purple line). Observe that the {feedforward} output based on the linear superposition model {exhibits} the largest RMS error at the 2nd layer (Fig. \ref{fig:NN_printing_result_sectn1}). {This is because this model only accounts for droplet deposition and does not capture any further dynamics.} The {feedforward} convRNN, with a control horizon of $Z=6$, {has} the best performance at this layer. {(Note that the \emph{learn \& control} strategy is solving a different MPC problem with $Z=3$)}. The {\emph{learn \& control}} algorithm takes the lowest RMS error at the 4th layer (Fig. \ref{fig:NN_printing_result_sectn2}) {because of feedback.} {Although} the {feedforward} convRNN profile is better laterally aligned with the reference than the {linear superposition model}, the volume of material being deposited is inadequate. By the final layer (Fig. \ref{fig:NN_printing_result_sectn3}), this volume inadequacy {results in} the greatest deviation from the reference. On the other hand, the \emph{learn \& control} feedback algorithm {improves} the RMS error over the convRNN-based {feedforward profile} by over $50\%$ (Fig. \ref{pic:N_part}) and the superposition-based {feedforward profile} by $45\%$. Using a 4.1 GHz Intel Core i7 16GB RAM computer, the computational time required for online model update and control calculations is about a half minute or less. Meanwhile, the typical print time for a layer is about 4 minutes. Since the computations and printing occur simultaneously, no additional lead time {is} required.


\emph{Discussion:} We have demonstrated that we can achieve significant improvement in reference geometry tracking in the inkjet 3D printing process using a feedback control scheme with minimal downtime. The height profile \textit{need not be measured frequently} and the computations required for model training and control learning, though expensive, may be carried out simultaneously with the printing process. These benefits are made possible {through} a physics-guided data-driven model that requires little data for training and is a good predictor of the process dynamics. {For larger grid sizes, the computational time will grow exponentially while the actual printing time grows linearly \cite{Inyang-Udoh2020}. Hence, future work will address decentralization of the MPC scheme to scale it almost linearly with grid size.}

\section{Conclusions} \label{sec:conclusion}

In this paper we proposed a novel predictive control method to improve geometry accuracy in jet-based AM. Our method improves upon existing linear MPC and iterative learning control methods by using a physics-guided data-driven model that captures the nonlinear fluid behavior of interacting droplets. We showed how the nonlinear predictive controller may be synthesized using backpropagation gradients. We also established conditions for stability of the controlled system. We implemented the feedforward nonlinear MPC scheme and showed it to outperform the state-of-the-art open-loop control for inkjet 3D printing. We {further} developed an efficient online learning and control algorithm that allows for feedback control in real time without adding substantial lead time to the fabrication process. The algorithm was also implemented on an inkjet 3D printing system and shown to substantially improve the reference geometry tracking over open-loop printing. { Future work will aim to distribute the MPC optimization for faster computation and implement the feedback control strategy in multi-material printing.}

\bibliographystyle{IEEEtran}
\bibliography{references3Dprint}
\end{document}